\documentstyle[11pt,amssymb,epsfig,amsmath]{article}

\begin{document}
\newcommand{\be}{\begin{equation}}
\newcommand{\ee}{\end{equation}}
\newcommand{\ba}{\begin{eqnarray}}
\newcommand{\ea}{\end{eqnarray}}
\newcommand{\no}{\nonumber \\}
\newcommand{\gsim}{\mathrel{\hbox{\rlap{\lower.55ex \hbox {$\sim$}}
                   \kern-.3em \raise.4ex \hbox{$>$}}}}
\newcommand{\lsim}{\mathrel{\hbox{\rlap{\lower.55ex \hbox {$\sim$}}
                   \kern-.3em \raise.4ex \hbox{$<$}}}}

\def\be{\begin{eqnarray}}
\def\ee{\end{eqnarray}}
\def\bea{\be}
\def\eea{\ee}
\newcommand{\e}{{\mbox{e}}}
\def\del{\partial}
\def\vr{{\vec r}}
\def\vk{{\vec k}}
\def\vq{{\vec q}}
\def\vp{{\vec p}}
\def\vP{{\vec P}}
\def\vt{{\vec \tau}}
\def\vs{{\vec \sigma}}
\def\vJ{{\vec J}}
\def\vB{{\vec B}}
\def\hatr{{\hat r}}
\def\hatk{{\hat k}}
\def\roughly#1{\mathrel{\raise.3ex\hbox{$#1$\kern-.75em%
\lower1ex\hbox{$\sim$}}}}
\def\lsim{\roughly<}
\def\gsim{\roughly>}
\def\fm{{\mbox{fm}}}
\def\vx{{\vec x}}
\def\vy{{\vec y}}
\def\({\left(}
\def\){\right)}
\def\[{\left[}
\def\]{\right]}
\def\EM{{\rm EM}}
\def\barp{{\bar p}}
\def\zz{{z \bar z}}
\def\mus{{\cal M}_s}
\def\abs#1{{\left| #1 \right|}}
\def\ve{{\vec \epsilon}}
\def\nlo#1{{\mbox{N$^{#1}$LO}}}
\def\MS{{\mbox{M1V}}}
\def\mut{{\mbox{M1S}}}
\def\Qt{{\mbox{E2S}}}
\def\rM{{\cal R}_{\rm M1}}\def\rE{{\cal R}_{\rm E2}}
\def\la{{\Big<}}
\def\ra{{\Big>}}
\def\lsim{\mathrel{\rlap{\lower3pt\hbox{\hskip1pt$\sim$}}
     \raise1pt\hbox{$<$}}} 
\def\gsim{\mathrel{\rlap{\lower3pt\hbox{\hskip1pt$\sim$}}
     \raise1pt\hbox{$>$}}} 
\def\N{${\cal N}\,\,$}

\def\ka{{\kappa}}
\def\lam{{\lambda}}
\def\dlt{{\delta}}
\def\lv{\lvert}
\def\rv{\rvert}

\def\J#1#2#3#4{ {#1} {\bf #2} (#4) {#3}. }
\def\PRL{Phys. Rev. Lett.}
\def\PL{Phys. Lett.}
\def\PLB{Phys. Lett. B}
\def\NP{Nucl. Phys.}
\def\NPA{Nucl. Phys. A}
\def\NPB{Nucl. Phys. B}
\def\PR{Phys. Rev.}
\def\PRC{Phys. Rev. C}

\renewcommand{\thefootnote}{\arabic{footnote}}
\setcounter{footnote}{0}

\vskip 0.4cm \hfill { }
 \hfill {\today} \vskip 1cm

\begin{center}
{\LARGE\bf Stress Tensor of Static Dipoles \\
 in strongly coupled $\cal{N}$=4  Gauge Theory
   }
\date{\today}

\vskip 1cm {\large Shu
Lin\footnote{E-mail:slin@grad.physics.sunysb.edu},
and Edward
Shuryak\footnote{E-mail:shuryak@tonic.physics.sunysb.edu}
 }


\end{center}

\vskip 0.5cm

\begin{center}

 {\it Department of Physics and Astronomy, SUNY Stony-Brook,
NY 11794}

\end{center}

\vskip 0.5cm

\begin{abstract}
     In the context of the AdS/CFT correspondence we calculate the
induced stress tensor of static dipoles
(electric-electric and electric-magnetic) in a strongly coupled ${\cal
N}=4$ SYM gauge theory, by  solving the linearized Einstein equation
with Maldecena string as a source. Analytic
expressions are given for the far-field and a near-field close to one
charge, and compared to what one has in weak coupling. 
The result can be compared to lattice results for QCD-like
theories in a deconfined but strongly coupled regime. 
\end{abstract}

\newpage

\renewcommand{\thefootnote}{\#\arabic{footnote}}
\setcounter{footnote}{0}

\section{Introduction }
AdS/CFT correspondence~\cite{adscft}  relates conformal
 $\cal N$=4 supersymmetric Yang-Mills theory (CFT)
 with  string theory in $AdS_5\times S_5$
space-time. Large number of colors
$N\rightarrow \infty$ and 't Hooft coupling
 $\lambda=g_{YM}^2N\rightarrow \infty$ further 
lead to the classical supergravity regime (weak coupling)
for the latter, putting the CFT into a  strong coupling regime.
In the decade since its invention, this correspondence became an
indispensable theoretical tool, providing
multiple interesting results about a strongly coupled regime
of  $\cal N$=4 supersymmetric gauge
theory. 

Among the earliest were
 calculation of the energy of
a static electric dipole  \cite{MALDA2_REY}, based on a shape of
``pending string'' held at the AdS boundary
at the positions of two static fundamental quarks, separated by
 distance $L$.  For further reference we will need the EOM of the
 string\footnote{Which is not the second order
equation
coming from the Lagrangian but the first (energy) integral of it.
The eqn(\ref{eq:shape}) represents half of the string, the other half
 is obtained by reflection $x\leftrightarrow -x$.}
\be \label{eq:shape}
x_z=-\frac{z^2}{\sqrt{z_m^4-z^4}}\ee
where $z_m$ is the maximal string
 extension into z direction.  We will also use notation
$L/2=x_m\approx0.60z_m$.

The resulting potential
\be 
E=-\frac{4\pi^2\(g_{YM}^2N\)^{1\over 2}}{\Gamma\(1\over{4}\)^4L}
\ee
has the famous factor $\sqrt{\lambda}$ (instead of $\lambda$ in
the weak coupling Coulomb law.
Soon this calculation was extended to include  magnetic objects
(monopoles and dyons)
by Minahan \cite{Minahan:1998xb}, which can be viewed
as an endpoints of the appropriate $D_1$
branes on the boundary. We will continue to discuss
puzzles related with the dipoles in the next subsection.

Naively one may interpret this answer by thinking
 of a strongly coupled vacuum
as a dielectric medium, with a dielectric constant
given by the ratio of the strong coupling result to
the zero order Coulomb\footnote{We remind the reader that
we include in it exchange due to
scalars. It is equal to that from gauge field
exchange  for quark-antiquark pair,
while for two quarks they have the opposite signs and cancel out.}:
 potential
\be \label{eqn_dielectric}
\epsilon={V_{Coulomb} \over V_{Maldacena}}={\sqrt{\lambda} 
\Gamma(1/4)^4  \over 8 \pi^3}\approx .636\sqrt{\lambda}\ee  
Although in a very qualitative sense this 
 idea is not wrong, it is certainly not literally true.
A proof of that are the calculations to be reported below,
which shows that
the stress tensor distribution in space is very different
from that in weak coupling. Of course, this is to be
expected, as the strongly coupled vacuum recieves nonperturbative 
modification from the
fields, leading to a nonlinear response.

Now, a decade later, there is a spike of activity
of using AdS/CFT to understand properties of
the deconfined phase of QCD, known as Quark-Gluon Plasma (QGP) 
\cite{Shu_QGP}. A number of phenomenological considerations lead to
a conjecture  \cite{SZ12} that QGP to be in a 'strongly coupled'
  regime
(sQGP) at temperatures not too high above the deconfinement
temperature $T=(1-2)T_c$. It is in this domain where RHIC experiments
at Brookhaven found a ``perfect liquid'' properties of sQGP.
Among  AdS/CFT-based works devoted to
it  are calculations of the energy loss
\cite{drag} and stress tensor imprint \cite{Gubser} of the moving objects
in thermal CFT plasma. Those are quite spectacular, providing in
particular a compete picture justifying another
hydrodynamical phenomenon, a ``conical flow'' in Mach direction
around the jet.

Lattice studies of sQGP  have also indicated features indicative
of a strong
coupling regime. Those most relevant for this work
obviously are studies of static charge pairs (electric or magnetic).
 Large deviation from a
perturbative picture of a screened Coulombic potential
are observed at $T$ above the deconfinement transition $T_c$.
More specifically, many features at $T>1.5Tc$ suggest a
``quasi-conformal'' regime, in which
all dimensional quantities (e.g. normalized 
energy density $\epsilon/T^4$) show weak $T$-dependence.
 
(Even larger deviations from perturbative approach -- the
Debye-screened charges -- are seen for static dipoles
at $T_c<T<1.5T_c$. Here the entropy and potential energy
associated with the string has very large part, linearly growing
with distance in some range, see e.g. \cite{Kaczmarek:2005zp}.
The question whether flux tubes --
remnants of confining strings -- can continue to exist
in a plasma phase was recently studied in \cite{Liao:2007mj}.)

More generally, the dynamics of electric and magnetic
gauge fields in a strongly coupled
 plasma remains very poorly understood. In particular, 
it has been suggested that sQGP contains large component
of magnetically charged quasiparticles -- monopoles and dyons, see
\cite{Liao_ES_mono,Chernodub_Zakharov}. Studies of the 
energy distribution in plasma induced by static dipoles
have been extensively done at zero temperature, demonstrating
existence of quantum confining string: unfortunately
similar calculations at $T>T_c$ are not yet available.
 One may wander whether the deconfined
QCD-like theories in those regimes are or are not
similar to the vacuum
of $\cal{N}$=4  Gauge Theory at strong coupling.

  These ideas motivated our present calculation, in which
we calculate stress tensor ``imprint'' of static dipoles in
 AdS/CFT. A simple diagrammatic picture of what is calculated is
provide by  Fig.\ref{fig_diagrams}.
Apart of the solutions themselves, to be given below in different 
regimes,
there are few particular issues  which would like to
investigate:\\
(i) How the dipole is seen at large distances $r\rightarrow \infty$?
What is the power of distance and its angular distribution?
Can it be related to expected behavior
of electric and scalar fields?
\\
(ii) What is the field near one of the charges? Can a non-singular
part corresponding to the fields of a second charge
and polarization cloud be identified? \\
(iii) Is there a visible remnant of the Matsubara
string, or a picture rather is of two
polarization clouds? In particular, what  is
the r.m.s. transverse size  $\sqrt{<y_{\perp}^2>}$
 at $\lv y\rv=0$ (the middle point)?\\
(iv) We will also consider an electric-magnetic pair: our main interest
in that is to see if there is some nontrivial
features related with electric-magnetic field interaction.

 \subsection{Strongly coupled versus weakly coupled dipoles} 

 One issue discussed in literature (after AdS/CFT potentials
been calculated) was whether some kind of diagram resummation can get
the reduction\footnote{We remind the reader that we discuss
  $\lambda\gg1$ regime.} of the coefficient, from $\sim\lambda$ to
$\sim\sqrt{\lambda}$. Semenoff and Zarembo\cite{Semenoff:2002kk}
have found that one can do so using ladder diagrams\footnote{Which is 
exact
 for a round
Wilson loop, approximate for rectangular ones.}. Shuryak and Zahed
\cite{Shuryak:2003ja} have noticed that such ladder diagrams
in a strongly coupled regime
imply a very short correlation time between colors of both
charges \be \delta t\sim L/\lambda^{1/4} \ee 
which will be crucial for understanding of the large-distance
field below.

The ``imprint'' of the pending string on the boundary
was first addressed by
Callan and Guijosa \cite{Callan:1999ki}, who
had calculated an ``image'' due to scalar (dilaton) field propagating
in the bulk. The boundary operator associated with a dilaton
is $trF^2$. Our work is very close to theirs, 
except that we calculate much more cumbersome graviton
propagation instead of a scalar one, 
to get the boundary stress tensor.

Their main results was a distribution of scalar density at large
distances from a dipole $r\gg L$ (i) has the form
\be trF^2(r)\sim L^3/r^7 \ee 
and (ii) is spherically symmetric.
Both are very different from 
what one finds for the shape
of the electric field of a weakly coupled electric dipole,
which has (i) power 6 and (ii) has a characteristic dipole
energy distribution $(3 cos^2(\theta)+1)$ ($\theta$ is polar angle
from a dipole direction). Our calculation to be reported also
will show power 7 but will have more complicated angular
distribution.

 The reason why the power is 7 rather than 6
was explained by Klebanov,Maldacena and Thorn \cite{Klebanov:2006jj}.
Imagine Euclidean time and perturbative diagram, in which
perturbative field of each charge can be written as a time integral
over a propagator, from a world line of a charge to an 
observation point: it produces power 6. The nontrivial point is that
in strongly coupled regime color time correlation \cite{Shuryak:2003ja}
mentioned above require both charges to emit quanta at 
the $same$ time; this changes a double 
time integral  into a single one,
  adding one more power of the distance.

\section{Solving the linearized Einstein equations in $AdS_5$}

As is clear from Introduction, the source of gravity in our problem
 are
strings extended into the AdS space. Naturally those are
considered to be weak sources, so 
 we will  linearize the 
 Einstein equations (with  $\Lambda=6$)
\be \label{eq:EE}
R_{\mu\nu}-{1\over{2}}Rg_{\mu\nu}+\Lambda g_{\mu\nu}=-\ka^2T_{\mu\nu} \ee
(\ref{eq:EE})
 and solve for small deviations
from the unperturbed
$AdS_5$ metric.

We choose to start with another form of (\ref{eq:EE}):
\be
R_{\mu\nu}+\(\Lambda-\frac{-\ka^2 T-\Lambda d}{2-d}g_{\mu\nu}\)=-\ka^2T_{\mu\nu}
\ee
with $d=5$
Linearizion of the above gives:
\be
\label{eq:lEE}
\dlt R_{\mu\nu}-4\dlt g_{\mu\nu}=\dlt S_{\mu\nu}
\ee
where $\dlt S_{\mu\nu}=-\ka^2\(\dlt T_{\mu\nu}-\frac{\dlt T}{3}g_{\mu\nu}\)$

We denote weak gravity perturbation as
 $\dlt g_{\mu\nu}=h_{\mu\nu}$ and use an axial gauge in which
the following components vanish
$h_{z\mu}=0\; (\mu=z,t,x^1,x^2,x^3)$. We use
 the usual Poincare coordinates for the AdS metric:
\be ds^2=\frac{-dt^2+d\vx^2+dz^2}{z^2} \ee
and  set the AdS radius $L_{ADS}$ to 1\footnote{Factors of $L_{ADS}$
 can be easily
reinstated by dimensional analysis.}.

Expressing the modifications of curvature
 $\dlt R_{\mu\nu}$ in terms of $h_{\mu\nu}$, 
(see Appendix.\ref{app:Rmunu} for a brief derivation)
we get the following equations

%
%
\ba
&&\label{eq:Rzz}{1\over 2}h_{,z,z}-{1\over {2z}}h_{,z}=\dlt S_{zz} \\
&&\label{eq:Rzm}{1\over 2}\(h_{,m}-h_m\)_{,z}=\dlt S_{zm} \\
&&\label{eq:Rmn}{1\over 2}\square h_{mn}-2h_{mn}+{z\over 2}h_{mn,z}
-{1\over 2}\(h_{m,n}+h_{n,m}\)+{1\over 2}\(h_{,m,n}-\Gamma_{mn}^z h_{,z}\)
=\dlt S_{mn}
\ea
where we have defined $h=g^{\lam\sigma}h_{\lam\sigma}$,
$h_m=g^{\lam\sigma}h_{\lam m,\sigma}$,
$\square=z^2\(-\del_t^2+\del_{\vx}^2+\del_z^2\)$, and from now
on latin indices stand for  4 boundary coordinates 
$(m,n=t,x^1,x^2,x^3)$.

We could in principle solve for $h$ from (\ref{eq:Rzz}), the 
result of which can help
to solve for $h_m$ from (\ref{eq:Rzm}). Finally solve for $h_{mn}$ with
$h$,$h_m$ plugged in (\ref{eq:Rmn}). However, we choose to do it
in a slightly different way:
As (\ref{eq:Rzm}) is first order in $z$, it is only a constraint equation.
With the boundary condition: $h_{mn}=0$(thus $h=0$,$h_m=0$) at $z=0$,
we obtain 
\be\label{eq:hm}
h_m=h_{,m}-2\int_0^z\dlt S_{zm}dz
\ee
(\ref{eq:Rzz}) is second order in $z$, but it gives also a constraint
when combined with (\ref{eq:Rmn}):
Denoting $\(m,n\)$ as the mn component of (\ref{eq:Rmn}),
$-\(t,t\)+\Sigma_i\(x^i,x^i\)$ gives:
\be\label{eq:hzz}
{1\over 2}h_{,z,z}-{7\over {2z}}h_{,z}=-\dlt S_{tt}+\Sigma_i \dlt S_{x^ix^i}
-2\int\(-\dlt S_{zt,t}+\Sigma_i \dlt S_{zx^i,x^i}\)dz
\ee

Combining (\ref{eq:Rzz}) and (\ref{eq:hzz}), we obtain the solution
for $h$
\be\label{eq:h}
h={1\over 3}\int_0^z dz\cdot z\(\dlt S_{zz}+\dlt S_{tt}-\Sigma_i \dlt S_{x^ix^i}
+2\int_0^z dz \(-\dlt S_{zt,t}+\Sigma_i \dlt S_{zx^i,x^i}\)\)
\ee

With $h$ obtained from (\ref{eq:h}) and $h_m$ eliminated, (\ref{eq:Rmn})
becomes a closed eqn for remaining components:
\be\label{eq:hmn}
{1\over 2}\square h_{mn}-2h_{mn}+{z\over 2}h_{mn,z}=s_{mn}
\ee
where a ``generalized source'' is
$s_{mn}=\dlt S_{mn}-\int_0^z\(\dlt S_{zm,n}+\dlt S_{zn,m}\)dz
+{1\over 2}h_{,m,n}+{1\over 2}\Gamma_{mn}^z h_{,z}$

The source terms created by the string are obtained from
the Nambu-Goto action
of the string in a standard way
\ba
&&S_{NG}=-\frac{1}{2\pi\alpha'}\int d^2\sigma\int d^5x
\sqrt{-det g}\dlt^{\(5\)}\(x-X\(\sigma\)\)      \no
&&\dlt T^{\mu\nu}=\frac{2\dlt S_{NG}}{\sqrt{-G}\dlt G_{\mu\nu}} \no
&&=\frac{-1}{\sqrt{-G}2\pi\alpha'}\int d^2\sigma\dlt^{\(5\)}\(x-X\(\sigma\)\)
\del_{\alpha}X^{\mu}\del_{\beta}X^{\nu}g^{\beta\alpha}
\ea
here we use $G_{\mu\nu}$ and $g_{\alpha\beta}$ to denote AdS metric and
induced metric respectively.

The string world sheet can be described by
\be
x^1=x\(t,z\),\: x^2=x^3=0
\ee

The resulting  source is  as follows (we use the order of coordinate
indices in the following 5-d matrices as $t,z,x^1,x^2,x^3$ and
all absent entries are zeros)
\be\label{eq:source}
&&\dlt S_{\mu\nu}=\frac{-\ka^2 z}{2\pi\alpha'}\dlt\(x^1-x\)\dlt\(x^2\)\dlt\(x^3\)
\frac{1}{\sqrt{1+x_z^2-x_t^2}} \no
&&\begin{pmatrix}
{{2x_t^2+x_z^2+1}\over 3}& x_tx_z& -x_t& & \\
x_tx_z& {{x_t^2+2x_z^2-1}\over 3}& -x_z& & \\
-x_t& -x_z& {{x_t^2-x_z^2+2}\over 3}& & \\
& & & -{2\over3}\(x_t^2-x_z^2-1\)& \\
& & & & -{2\over3}\(x_t^2-x_z^2-1\)
\end{pmatrix}
\ee

With (\ref{eq:hmn}),(\ref{eq:h}) and (\ref{eq:source}), we can 
 solve for $h_{mn}$, provided any explicit profile of the string. We will 
do this for three different string profiles 
separately in the following sections, and extract the corresponding
stress tensors.

\section{The stress tensor  of a static quark}

As a warm up, we will start with the case of a straight string, which
corresponds to a single quark in $\cal N$=4 SYM. The string profile
is simply $x\(t,z\)=0$. Substitute in (\ref{eq:source}), we obtain:
\be\label{eq:sq_source}
\dlt S_{\mu\nu}=\frac{-\ka^2 z}{2\pi\alpha'}\dlt\(x^1\)\dlt\(x^2\)\dlt\(x^3\)
\begin{pmatrix}
{1\over 3}& & & & \\
& -{1\over 3}& & & \\
& & {2\over 3}& & \\
& & & {2\over3}& \\
& & & & {2\over3}
\end{pmatrix}
\ee

Static source leads to the metric perturbation
$h_{mn}$ which is time-independent. Performing a Fourier transform 
$h_{mn}^k=\int h_{mn}e^{i\vk\vx}d^3x$
 we convert the PDE (\ref{eq:hmn}) to an ODE:
\be\label{eq:fthmn}
{1\over 2}z^2\(h_{mn,z,z}^k-k^2h_{mn}^k\)-2h_{mn}^k+{z\over 2}h_{mn,z}^k=s_{mn}^k
\ee
An upper index k will be used 
below to indicate a Fourier transformed quantity.

$S_{\mu\nu}^k$ is just $S_{\mu\nu}$ without delta functions.
$h^k$ and $s_{\mu\nu}^k$ have simple forms displayed as follows:
\ba
&&h^k=-{2\over 9}\frac{-\ka^2}{2\pi\alpha'}z^3\\
&&s_{mn}^k=\frac{-\ka^2}{2\pi\alpha'}
\[
\begin{pmatrix}
{2\over 3}& & & \\
& {1\over 3}& & \\
& & {1\over3}& \\
& & & {1\over3}&
\end{pmatrix}z+
\begin{pmatrix}
0& & & \\
& & & \\
& & k_mk_n& \\
& & & 
\end{pmatrix}{z^3\over 9}
\]
\ea
The equation is Bessel type and can be dealt with using 
a Green function built out of such functions. Instead
 we consider
 a more general equation with arbitrary power of $z$ in the source
\be\label{eq:zn}
{1\over 2}z^2\(h_{mn,z,z}^k-k^2h_{mn}^k\)-2h_{mn}^k+{z\over 2}h_{mn,z}^k=c_nz^n
\ee
%
which is directly
 solvable in terms of Meijer-G funcion and hypergeometric
function:
\be\label{eq:mg}
h_{mn}^k=I_2(kz)\(C_2+G_{1,3}^{2,1}\(\frac{k^2z^2}{4}\Big\vert^{1}_{{n\over 2}+1,{n\over2}-1,0}\)\frac{2^{n-1}}{k^n}\) \no
+K_2(kz)\(C_1-_1F_2\(^{{n\over 2}+1}_{{n\over 2}+2,3}\Big\vert\frac{k^2z^2}{4}\)
{\frac{k^2z^{n+2}}{4n+8}}\)
\ee

The constants $C_1$ and $C_2$ are to be fixed by boundary conditions.
One of the condition is the metric perturbation vanishes at AdS boundary,
i.e. $h_{mn}^k=0$ at $z=0$, which fixes $C_1=0$. The other boundary condition
proposed in \cite{starinets} for thermal AdS is incoming
metric perturbation at the horizon. However in our case, we need
a different boundary condition due to the absence of horizon in AdS.
Since $h^k$ grows as $z^3$ in the present case, while $h_{mn}^k$
show possible exponential growth at large z. It is natural to
 propose no exponential growth at $z=\infty$ as the boundary condition.

At large z, only the first term containing $I_2(kz)$ is dominant,
the boundary condition becomes:
\ba
C_2+\frac{2^{n-1}}{k^n}
G_{1,3}^{2,1}\(\frac{k^2z^2}{4}\Big\vert^{1}_{{n\over 2}+1,{n\over2}-1,0}\)=0 \no
\ea
The asymptotic of Meijer-G function ($z\rightarrow \infty$) gives:
\ba
G_{1,3}^{2,1}\(\frac{k^2z^2}{4}\Big\vert^{1}_{{n\over 2}+1,{n\over2}-1,0}\)
\rightarrow \Gamma\({n\over 2}+1\)\Gamma\({n\over 2}-1\) \no
\ea
 which finally fixes $C_2=-\frac{c_n2^{n-1}}{k^n}
\Gamma\({n\over 2}+1\)\Gamma\({n\over 2}-1\)$.
Applying it to our source
 $s_{mn}=c_1z+c_3z^3$, where $c_1$ and $c_3$ are matrix-valued
(the indices are suppressed here), we have 
$C_2={\pi\over k}c_1-{3\pi\over k^3}c_3$

The stress tensor of the corresponding boundary CFT is proportional to
the coefficient of $z^2$ term, \footnote{We remind the reader
that unperturbed metric has $1/z^2$ and thus the relative smallness
is $O(z^4)$ fitting the dimension of the stress tensor.} 
 which we denote as $Q_{mn}$ throughout 
this paper, in small z expansion of $h_{mn}$. 
The precise relation can be obtained from (36) of \cite{Gubser}, which in
our case is simply (with $L_{ADS}=1$):
\be
T_{mn}=\frac{2}{\ka^2}Q_{mn}
\ee

Note $G_{1,3}^{2,1}\(\frac{k^2z^2}{4}\Big\vert^{1}_{{n\over 2}+1,{n\over2}-1,0}\)$ and
$_1F_2\(^{{n\over 2}+1}_{{n\over 2}+2,3}\Big\vert\frac{k^2z^2}{4}\)z^{n+2}$ contains
only odd power of z for odd $n$, thus does not contribute to $Q_{mn}$.
We have
\be
Q_{mn}={1\over8}k^2C_2
\ee

Reinstate the factor $L_{ADS}^2$, together with the relation 
$\frac{L_{ADS}^2}{\alpha'}=\sqrt{\lam}$, we have the final stress tensor:

\be
T_{mn}^k=\frac{-\sqrt{\lam}}{\pi}\frac{k^2}{8}
\[
\begin{pmatrix}
{2\over 3}& & & \\
& {1\over 3}& & \\
& & {1\over3}& \\
& & & {1\over3}
\end{pmatrix}{\pi\over k}-
\begin{pmatrix}
0& & & \\
& & & \\
& & k_mk_n& \\
& & & 
\end{pmatrix}{\pi\over {3k^3}}
\]
\ee

It is easy to verify the stress tensor above is traceless $T_{mn}\eta^{mn}=0$,
which is a consequence of conformal invariance. It also satisfies
the conservation of energy and momentum $k^mT_{mn}=0$.
In doing inverse Fourier transform, we find the k-integrals are not 
well-defined. One trick is to introduce a regulator $e^{-ak}\(a>0\)$
to the integral, and take the limit $a\rightarrow 0$ in the final answer.
We end up with the following result:

\be\label{eq:sq_st}
T_{mn}=\frac{-\sqrt{\lam}}{\pi}\frac{1}{8\pi}
\[-{2\over{3r^4}}
\begin{pmatrix}
1& & & \\
& 1& & \\
& & 1& \\
& & & 1
\end{pmatrix}+
\begin{pmatrix}
0& & & \\
& & & \\
& & y_my_n& \\
& & & 
\end{pmatrix}{4\over {3r^6}}
\]
\ee
where r is the distance from the quark.  
The ${1\over r^4}$ power is obvious by dimension.
Let us recall the result obtained in \cite{Callan:1999ki}\footnote
{There is a typo in eqn (23) of the paper. We quote the corrected expression}
\be\label{eq:trF2}
&&{\cal O}_{F^2}={1\over{4g_{YM}^2}}trF^2+\cdots
={1\over{2g_{YM}^2}}tr(-E^2+B^2)+\cdots \no
&&={1\over{32\pi^2}}\frac{\sqrt{\lam}}{r^4}
\ee
While in our case, the $T_{00}$ component gives
\be\label{eq:T00}
&&T_{00}={1\over{2g_{YM}^2}}tr(E^2+B^2)+\cdots \no
&&={1\over{12\pi^2}}\frac{\sqrt{\lam}}{r^4}
\ee

In both (\ref{eq:trF2}) and (\ref{eq:T00}), the dots represent contributions
from scalars and gluinoes. 
 If we assume the magnetic field is not present,
the difference in the two operators implies significant contribution
are received from the scalars and gluinoes.

\section{The stress tensor image of static electric dipole}

Now we turn to the Maldacena's pending string, the ends of which
attached to a quark and antiquark, corresponding to
a static electric dipole. The string profile $x(z)$ is double-valued.
We use $\pm x(z)\:(x(z)>0)$ to denote two halves of the string. 
The EOM can be integrated to give $x(z)$
 in terms of elliptic integrals. We will
not refer to explicit form until the end of the calculation.

The source term and its Fourier transformed version
are a bit complicated:
\be\label{eq:dp_source}
&&\dlt S_{\mu\nu}=\frac{-\ka^2 z}{2\pi\alpha'}\dlt\(x^1-x(z)\)\dlt\(x^2\)\dlt\(x^3\)
\frac{1}{\sqrt{1+x_z^2}}\no
&&\begin{pmatrix}
{{x_z^2+1}\over 3}& & & & \\
& {{2x_z^2-1}\over 3}& -x_z& & \\
& -x_z& {{-x_z^2+2}\over 3}& & \\
& & & {{2x_z^2+2}\over3}& \\
& & & & {{2x_z^2+2}\over3}
\end{pmatrix}
+(x\rightarrow -x)  \ee
\be
&&\dlt S_{\mu\nu}^k=\frac{-\ka^2 z}{2\pi\alpha'}
\frac{2}{\sqrt{1+x_z^2}}
\biggl[
\begin{pmatrix}
{{x_z^2+1}\over 3}& & & & \\
& {{2x_z^2-1}\over 3}& & & \\
& & {{-x_z^2+2}\over 3}& & \\
& & & {{2x_z^2+2}\over3}& \\
& & & & {{2x_z^2+2}\over3}
\end{pmatrix}\cos(k_1x) \no
&&+\begin{pmatrix}
0& 0& 0& & \\
0& 0& -x_z& & \\
0& -x_z& 0& & \\
& & & 0& 0\\
& & & 0& 0
\end{pmatrix}i\sin(k_1x)
\biggr]
\ee

It is understood that the source term vanishes for $z>z_m$. (\ref{eq:h})
and (\ref{eq:hmn}) gives:
\ba\label{eq:smn}
&&h(z<z_m)={1\over 3}\frac{-\ka^2}{2\pi\alpha'}F(z)\;   \\ \nonumber
&&h(z>z_m)={1\over 3}\frac{-\ka^2}{2\pi\alpha'}\(F(z_m)+{1\over 2}(z^2-z_m^2)G(z_m)\)
\;   \ea

\ba
&&s_{mn}^k(z<z_m)=\frac{-\ka^2}{2\pi\alpha'}
\biggl[
{1\over 3}E_1(z)
\begin{pmatrix}
1& & & \\
& -1& & \\
& & 2& \\
& & & 2
\end{pmatrix}
+{1\over 3}E_2(z)
\begin{pmatrix}
1& & & \\
& -1& & \\
& & 2& \\
& & & 2
\end{pmatrix} \no 
&&+H(z)
\begin{pmatrix}
0& & & \\
& 2k_1& k_2& k_3\\
& k_2& & \\
& k_3& &
\end{pmatrix} 
-{1\over3}F(z)
\begin{pmatrix}
0& & & \\
& & & \\
& & \frac{k_mk_n}{2}& \\
& & & 
\end{pmatrix}
+\frac{F'(z)}{6z}
\begin{pmatrix}
-1& & & \\
& 1 & & \\
& & 1& \\
& & & 1
\end{pmatrix}
\biggr]
\;  \\
&&s_{mn}^k(z>z_m)=\frac{-\ka^2}{2\pi\alpha'}
\biggl[
H(z_m)
\begin{pmatrix}
0& & & \\
& 2k_1& k_2& k_3\\
& k_2& & \\
& k_3& & 
\end{pmatrix}
-{1\over3}\(F(z_m)+{1\over2}(z^2-z_m^2)G(z_m)\) \no
&&\begin{pmatrix}
0& & & \\
& & & \\
& & \frac{k_mk_n}{2}& \\
& & & 
\end{pmatrix}
+\frac{G(z_m)}{6}
\begin{pmatrix}
-1& & & \\
& 1& & \\
& & 1& \\
& & & 1
\end{pmatrix}
\biggr] \;  
\ea

with
\ba
&&E_1(z)=\frac{2z}{\sqrt{1+x_z^2}}x_z^2\cos(k_1x) \no
&&E_2(z)=\frac{2z}{\sqrt{1+x_z^2}}\cos(k_1x) \no
&&F(z)=\int_0^z\frac{-4\cos(k_1x)}{\sqrt{1+x_z^2}}z^2dz+
\int_0^z dz\(z\int_0^z\frac{-4\sin(k_1x)}{\sqrt{1+x_z^2}}zk_1x_zdz\) \no
&&G(z_m)=\int_0^{z_m}\frac{-4\sin(k_1x)}{\sqrt{1+x_z^2}}zk_1x_zdz \no
&&H(z)=\int_0^z\frac{2z\sin(k_1x)}{\sqrt{1+x_z^2}}x_zdz
\ea

With the explicit expression of $s_{mn}^k$, we can build the general
solution to (\ref{eq:hmn}):
\be\label{eq:dp_hmn}
h_{mn}^k=I_2(kz)C_2+K_2(kz)C_1
+2\(I_2(kz)\int_{z_m}^z\frac{s_{mn}^k(z)K_2(kz)}{z}dz
-K_2(kz)\int_{z_m}^z\frac{s_{mn}^k(z)I_2(kz)}{z}dz\)
\ee

At large z, no exponential growth condition requires
$C_2+2\int_{z_m}^\infty\frac{s_{mn}^k(z)K_2(kz)}{z}dz=0$. 
The convergence of the integral is ensured by $K_2(kz)$ in the
integrand . At small z, $s_{mn}^k\sim O(z)$ while $I_2(kz)\sim O(z^2)$
the integral containing $I_2(kz)$ is finite as z approach 0, therefore
the boundary condition gives: $C_1-2\int_{z_m}^0\frac{s_{mn}^k(z)I_2(kz)}{z}dz=0$.

In order to extract the stress tensor, we need to collect $z^2$ terms.
 It is helpful to  write down the series expansion of the two integrals
\be
\int_{z_m}^z\frac{s_{mn}^k(z)I_2(kz)}{z}dz=a_0+a_1z+\cdots \\
\int_{z_m}^z\frac{s_{mn}^k(z)K_2(kz)}{z}dz={b_{-1}\over z}+b_0+\cdots
\ee

The coefficient of $z^2$ is given by:
$Q_{mn}={1\over8}k^2C_2+{k^2\over4}b_0$. 
Note $C_1$ does not appear in the expression
 We may also write it as
\be\label{eq:k_st}
Q_{mn}^k=-{k^2\over4}\int_{z_m}^\infty\frac{s_{mn}^k(z)K_2(kz)}{z}dz
+{k^2\over4}\lim_{{\epsilon\rightarrow0}}
\(\int_{z_m}^\epsilon\frac{s_{mn}^kK_2(kz)}{z}dz-{b_{-1}\over\epsilon}\)
\ee

We could proceed in momentum space. However it turns out to be
 much easier and illustrating
to do inverse Fourier transform and continue in configuration space
from now on. 

A nice property of Fourier transform is
 ${\cal F}^{-1}(F(k)G(k))=\int f(x)g(y-x)dx$. Identifying the source
dependent $s_{mn}^k$ as $F(k)$, the inverse Fourier transform 
of which gives $f(x)$. Correspondingly, each $G(k)$ is transformed
to $g(y-x)$. The latter can be interpreted as a propagator from
a point on the source $x$ to a point on the boundary $y$. With
this in mind, we define the following propagator:

\ba\label{eq:prop}
&&P_s(\vy-\vx)=\frac{1}{(2\pi)^3}\int k^2K_2(kz)e^{-i\vk\vx}d^3k
={15\over{4\pi}}\frac{z^2}{\(z^2+r^2\)^{7\over2}}
\ea

Let us take a moment to worry about the term involving $b_{-1}$. By analyzing
small z behavior of $s_{mn}$ and $I_2(kz)$, we find $b_{-1}=\frac{\#}{k^2}$.
Inverse Fourier transform of ${k^2\over4}b_{-1}$ is not well-defined.
Again we introduce the same regulator $e^{-ak}$ as in the previous section.
We find a vanishing result after taking the limit $a\rightarrow0$
\footnote{this may seems problematic. Actually the same regularization
can also be applied to $K_2(kz)$ if we first expand it in series of k.
The non-vanishing terms match those obtained from 
series expansion of propagator $P_s$ in r}

Finally, we can write the stress tensor in a very short form:
\be\label{eq:dp_st2}
&&Q_{mn}=-{1\over4}\int_{z_m}^\infty dz
\int \frac{s_{mn}(z,\vx)P_s(\vy-\vx)}{z}d^3x \no
&&+{1\over4}\int_{z_m}^0dz\int \frac{s_{mn}(z,\vx)P_s(\vy-\vx)}{z}d^3x
\ee

Before proceeding with the calculation, we would like to make 
few general
comments: (i) The trace of the stress tensor is  given by the coefficient
of $z^4$ term of $h$. From (\ref{eq:smn}), we find
that $h\sim F(z)$ at small
z does $not$ contain $z^4$ term, therefore we expect the final stress
tensor to be $traceless$, which is also required by conformal invariance.
(ii) The divergence of the stress tensor $\del_{\lam}T_{\lam m}$ turns out to be the
the coefficient of $z^4$ term of $h_m$. From (\ref{eq:hm}) and (\ref{eq:smn})
we conclude $\del_{\lam}T_{\lam m}=\frac{\sqrt{\lam}}{\pi}{1\over{2z_m^2}}
\(\dlt(x^1-x_m)-\dlt(x^1+x_m)\)\dlt(x^2)\dlt(x^3)\dlt_{m1}$. 
The divergence is non-vanishing only for $m=x^1$ at 
the end points of the string where
 the quark and antiquark are placed. It corresponds to a pair of 
antiparallel forces
 which hold  quark and antiquark, preventing them
from falling onto each other. This will be another
general condition to be
satisfied by the stress tensor.

\subsection{Far field}

With (\ref{eq:dp_st2}) at hand, we first calculate the stress tensor
in region far from the dipole. The inverse Fourier transforms of 
$s_{mn}$ are linear combinations of those of $E_1,E_2,F,G,H$.
Such terms as $k_mH$  can be replaced  by
$i{\del_{x_m}}H=-i{{\overleftarrow\del}_{x_m}}H=i{{\overleftarrow\del}_{y_m}}$.
In the first identity, we use
partial integration so that the derivative only acts on the propagators
(we indicate this with a left arrow on top of the derivative). The second 
identity is due to $P_s=P_s(\vy-\vx)$.
Similarly, $k_mk_nF\rightarrow 
-{\overleftarrow\del}_{x_m}{\overleftarrow\del}_{x_n}F\rightarrow
-{\overleftarrow\del}_{y_m}{\overleftarrow\del}_{y_n}F$

We list the back-transformed result of $E_1,E_2,F,G,H$ here:
\ba\label{eq:E-H}
&&E_1=\frac{z^5}{z_m^2\sqrt{z_m^4-z^4}}\dlt(x^2)\dlt(x^3)\dlt(x^1-x(z))
+(x^1\rightarrow-x^1) \\
&&E_2=\frac{z\sqrt{z_m^4-z^4}}{z_m^2}\dlt(x^2)\dlt(x^3)\dlt(x^1-x(z))
+(x^1\rightarrow-x^1) \\
&&F={4\over {z_m^2}}\dlt(x^2)\dlt(x^3)
\frac{-(z^2-z_1^2)(z_m^4-3z_1^4)}{4z_1^2}\theta(z-z_1)+(x^1\rightarrow-x^1) \\
&&G={4\over{z_m^2}}\dlt(x^2)\dlt(x^3)
\(-\frac{z_m^4-3z_1^4}{2z_1^2}\theta(z-z_1)
+\frac{z_m^4-z_1^4}{2z_1}\dlt(z-z_1)\)+(x^1\rightarrow-x^1)\\
&&H=-{1\over{z_m^2}}\dlt(x^2)\dlt(x^3)\frac{z_1\sqrt{z_m^4-z_1^4}}{i}
\theta(z-z_1)-(x^1\rightarrow-x^1)
\ea

with $z_1=z_1(x^1)\,(0<x^1<x_m)$, the inverse function of $x(z)$. Contribution
from negative $x^1$ is included in the second term for each function.
Note $E_1,E_2,F,G$ are symmetric in $x^1$, while $H$ is antisymmetric.

In order to obtain the far field stress tensor, we need to perform
a large $\lv y\rv$ expansion of the stress tensor. Note the y-dependence
enters the stress tensor via the propagator, we can do a large $\lv y\rv$
expansion on the propagator in the second term since $z<z_m\ll\lv y\rv$.
While for the first integral, z extending to infinity, we need to
do the integral first before a valid expansion is possible. Fortunately
this time the source has very simple z-dependence: $s_{mn}(z)=\#+\#z^2$.
The rest of the calculation is straight forward. After collecting all
terms, we find the first nontrivial result appears at the order
${1\over{\lv y\rv^7}}$. The power again agree with the result of $trF^2$ 
obtained in \cite{Callan:1999ki}.
 We list the stress tensor as follows
(up to the order ${1\over{\lv y\rv^7}}$):

\ba\label{eq:ff}
&&T_{00}={1\over4}\frac{-\sqrt{\lam}}{\pi}\frac{15}{4\pi}
\(\frac{a_G(7y_1^2-y^2)}{12\lv y\rv^9}
+\({a_{E1}\over3}+{a_{E2}\over3}+{a_{F}\over6}\){1\over{\lv y\rv^7}}\) \\
&&T_{0m}=0 \\
&&T_{mn}={1\over4}\frac{-\sqrt{\lam}}{\pi}\frac{15}{4\pi}
\biggl[
\(-7a_H+{7\over6}a_G\)
\begin{pmatrix}
2y_1& y_2& y_3\\
y_2& & \\
y_3& & 
\end{pmatrix}
\frac{y_1}{\lv y\rv^9}
+{2\over3}\(a_{E1}+a_{E2}\)\dlt_{mn}{1\over{\lv y\rv^7}} \no
&&-\(a_{E1}+{a_G\over6}-2a_H\)
\begin{pmatrix}
1& 0& 0\\
0& & \\
0& & 
\end{pmatrix}
{1\over{\lv y\rv^7}}
-\({7a_F\over6}-{7a_G\over12}\)\frac{y_my_n}{\lv y\rv^9}
-{21a_G\over4}\frac{y_my_ny_1^2}{\lv y\rv^{11}}
\biggr]
\ea
with
\ba
&&a_G=2\int_0^{x_m}G(z_m)\(x^1\)^2dx^1=-0.7189z_m^3 \no
&&a_F=2\int_0^{x_m}F(z_m)dx^1=-0.9585z_m^3 \no
&&a_H=2\int_0^{x_m}H(z_m)x^1dx^1=0.1797z_m^3 \no
&&a_{E1}=\int_{z_m}^0E_1(z)dz=-0.7189z_m^3 \no
&&a_{E2}=\int_{z_m}^0E_2(z)dz=-0.4793z_m^3 \no
\ea

We can verifiy explicitly that the stress tensor is traceless and
divergence-free at this order.

  Now we proceed to analysis of the results, describing which
features are general and should be expected and which 
of them are qualitatively new.

A vanishing energy flux (Poynting vector) $T_{0m}=0$ 
is related with zero magnetic field expected for static electric
configuration. Indeed, a time reversal would change the sign of
the magnetic field and 
the Poynting vector, but leaves the problem invariant.

Having said that, we by no means imply that the only field
in question is the electric field. Indeed, vacuum polarization
should include all other fields of the theory, and perturbatively
we know that all color fields  of the theory --
 gluinoes and scalars -- should contribute, to charge polarization
density as well as to the energy we calculate. However,
a very simplistic view of the scalars\footnote{Ignoring quartic terms with
commutators of various flavor components.} based on $(\partial_\mu
\phi)^2$ Lagrangian would produce the same distributions as a
vector field, since that can be viewed as just generated by another
scalar field $A_0$.

  The obvious point of comparison is stress tensor distribution
for a perturbative dipole. Its electric field
\be E_m(y) =({g^2\over 4\pi})\left({y_m-(L/2)e_m \over |y_m-(L/2)e_m|^3}-{y_m+(L/2)e_m \over |y_m+(L/2)e_m|^3}\right) \ee
leads to stress tensor which is at large distances $\sim L^2/y^6$.
 The result we obtain is $\sim L^3/r^7$: the difference is due to the
a phenomenon of
``short-time-color-locking'' \cite{Shuryak:2003ja,Klebanov:2006jj}
we already
discussed in the Introduction. Perhaps another way to
explain it is to say that a scalar density, induced by a
dipole, is large in all the volume $\sim L^3$. 

\begin{figure}[t]
  \epsfig{file=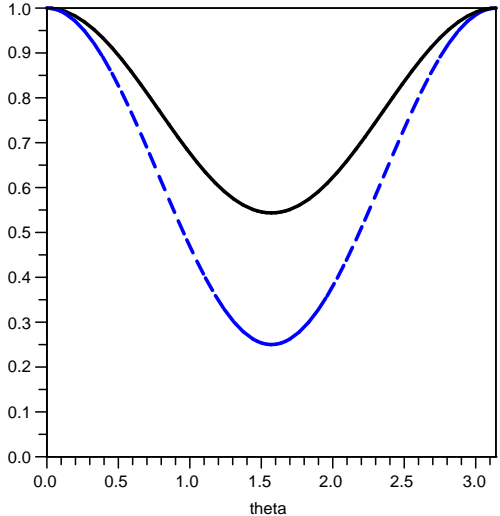,width=8.cm}
  \vspace{0.1in} \caption{(Color online) The  far field energy
distribution in polar angle $\theta (cos(\theta)=y_1/\lv y\rv)$,
normalized at zero angle. Solid (black) line is our result,
  compared to the perturbative result $(3cos^2+1)/4$ given by
the dashed (blue) line. } \label{fig_energy}
\end{figure}

  Let us now
comment on the angular distribution. Perturbative dipole field
at large distances contains the first power of the dipole vector:
thus its angular momentum is 1. Energy density constructed out
of this field, obviously has only angular momenta 2 and 0,
or powers of $cos^n(\theta)$ with $n=0,2$.  Stress tensor also
contains such
components, but also terms of the  type 
$T_{mn}\sim y_my_n(\vec L \vec y)^2$. Looking at our result we 
find that indeed {\em no other} angular structures appeared.
This is to be expected, as electric field is still the
only vector field of the theory. 
The angular distribution of the far field energy
is compared to the perturbative result in Fig.\ref{fig_energy}:
although there is tendency to a more spherical distribution
(like obtained for scalar density \protect\cite{Callan:1999ki}),
the peaks in the dipole directions are still there.

One more simple case to discuss is the stress tensor
on a line connecting the charges:
by symmetry transverse component of the field $\vec E_\perp=0$
and only $E_x$ remains. The Maxwellian tensor then should 
satisfy $T_{22}=T_{33}=-T_{11}=T_{00}$: and the result we obtain
does not satisfy it.
 We thus see once again, that gluino and scalar parts
of the stress tensor $must$ contribute 
to the far field asymptotic in question. 


\subsection{A field near one charge}

Next we would like to study the stress tensor near one of the charge.
For this purpose, we make a shift of variables
$y_1\rightarrow y_1+x_m,\,y_2\rightarrow y_2,\,y_3\rightarrow y_3$, and
consider small $\lv y\rv$ behavior of the stress tensor.

It is clear from the single charge result(\ref{eq:sq_st}) the stress
 tensor will blow up as $\lv y\rv\rightarrow0$, so to the
leading order in $\lv y\rv$, we may focus on
 its divergent part only.

Let us recall the basic expression for the stress tensor:
\be
&&Q_{mn}=-{1\over4}\int_{z_m}^\infty dz
\int \frac{s_{mn}(z,\vx)P_s(\vy-\vx)}{z}d^3x \no
&&+{1\over4}\int_{z_m}^0dz\int \frac{s_{mn}(z,\vx)P_s(\vy-\vx)}{z}d^3x
\ee

As $\lv y\rv\rightarrow0$, the first
term is finite($z>z_m$), which we ignore as discussed above.
While the propagator in the second term
$P_s={15\over{4\pi}}\frac{z^2}{\(z^2+r^2\)^{7\over2}}$ contains a
singularity at $z=0,\,r=0$, which leads to a possible divergence
in stress tensor(unless the source provide enough powers of z).
We can also claim the divergence is from integration at small z.
 Since the integral involving $s_{mn}$ and $P_s$
cannot be done analytically, a careful analysis is needed
to obtain the leading terms in Laurent expansion of the stress tensor.

We first use the common factor $\dlt(x^2)\dlt(x^3)$
in the source to simplify the propagator:
\be
&&r^2=(y_1+x_m-x^1)^2+(y_2-x^2)^2+(y_3-x^3)^2 
=(y_1-\Delta x)^2+y_2^2+y_3^2 \no
&&=r_0^2-2y_1\Delta x+\Delta x^2 \nonumber
\ee
with $r_0^2=y_1^2+y_2^2+y_3^2$, $\Delta x=x^1-x_m$. Then
the propagator can be expanded in $\Delta x^1$:
\be
P_s={15\over{4\pi}}\(\frac{z^2}{\(z^2+r_0^2\)^{7\over2}}
-7\frac{z^3}{\(z^2+r_0^2\)^{9\over2}}y_1\Delta x^1+\cdots\)
\ee
Note the leading term of the propagator does not 
depend on $x^1,x^2,x^3$.
A similar trick is used as in the case of far field:
$k_m=-i{\overleftarrow\del}_{x_m}=i{\overleftarrow\del}_{y_m}$.
The second identity is due to $P_s=P_s(\vy-\vx)$.
 If only the leading order
result of the stress tensor is needed,
we perform the x-integral with the source,
 keeping the smallest power in z(As we argued before
smaller power of z corresponds to larger term in
expansion of the stress tensor):
\ba
&&\int E_1d^3x \sim\frac{z^5}{z_m^4} \no
&&\int E_2d^3x \sim z \no
&&\int Fd^3x \sim -z_m^2\int(z^2-z_1^2)\theta(z-z_1)dx^1
=-{2\over3}z^3 \no
&&\int {F'\over z}d^3x \sim -2z_m^2\frac{z}{z_1^2}\theta(z-z_1)dx^1
=-2z \no
&&\int -iHd^3x \sim z_1\theta(z-z_1)dx^1
=\frac{z^4}{4z_m^2} \nonumber
\ea

Convolute the above results with the leading order propagator,
we find they give the following divergence:
\ba
&&E_1\rightarrow ln(r_0) \no
&&E_2,{F'\over z}\rightarrow {1\over r_0^4} \no
&&F\rightarrow {1\over r_0^2} \no
&&iH \rightarrow {1\over r_0} \nonumber
\ea

Therefore the leading order result is given by
$E_2,{F'\over z}$ and $F$. The last also give ${1\over r_0^4}$
when combined with the double
derivatives in the coefficient. Collecting all the contributions,
we find the leading near field contribution, which is of course
precisely the stress tensor of a single charge (\ref{eq:sq_st})
\be\label{eq:nf1}
T_{mn}^{LO}=\frac{-\sqrt{\lam}}{\pi}\frac{1}{8\pi}
\biggl[
-{2\over{3r_0^4}}
\begin{pmatrix}
1& & & \\
& 1& & \\
& & 1& \\
& & & 1
\end{pmatrix}+
\begin{pmatrix}
0& & & \\
& & & \\
& & y_my_n& \\
& & & 
\end{pmatrix}{4\over {3r_0^6}}
\biggr]
\ee

The aim now is to extend the analysis to the next order
correction to (\ref{eq:nf1}). Note the correction from the source
will give at least $O(z^4)$ correction, while that from the
propagator is of $O(\Delta x^1)\sim z_1^3\sim z^3$, with an
additional $z^2+r_0^2$ in the denominator. As a result,
we can keep the leading order source but care about the correction
from the propagator when necessary. Finally we find the
next order correction to the stress tensor is from the LO source 
$E_2,F,{F'\over z}$ convoluted with the
 NLO correction to the propagator
$-7\frac{z^3}{(z^2+r_0^2)^{9\over2}}y_1\Delta x^1$, as well as the
leading result from $iH$. We display the correction to the
near field as follows:
\be\label{eq:nf2}
&&T_{{mn}}^{NLO}=\frac{-\sqrt{\lam}}{\pi}\frac{1}{12\pi}\frac{1}{z_m^2}
\biggl[
{y_1\over{6r_0^3}}
\begin{pmatrix}
5& & & \\
& 8& & \\
& & 8& \\
& & & 8
\end{pmatrix}-
\begin{pmatrix}
0& & & \\
& 2y_1& y_2& y_3\\
& y_2& & \\
& y_3& & 
\end{pmatrix}\frac{4}{3r_0^3} \no
&&-\begin{pmatrix}
0& & & \\
& & & \\
& & y_my_n& \\
& & & 
\end{pmatrix}\frac{y_1}{2r_0^5}
\biggr]
\ee

We can also verify the stress tensor at this order is traceless and
divergence-free.

  Let us now analyze the results and compare it with expectations.
In general one can expect that close to the charge there is
a singular electric field $E^{sing}\sim 1/r_0^2$ 
plus a finite field induced by all other
charges. 
\be E_iE_j\approx E^{sing}_i E_j^{sing}+ E^{sing}_i E_j^{reg}+ E^{sing}_j E_i^{reg}+... \ee
The scalar field in weak coupling add the same distributions.

 If the vacuum would be a simple dielectric,
both the singular and regular field would be just
free fields times the dielectric constant (\ref{eqn_dielectric}),
and the relative correction be the same. Let us see whether this idea
works or not.
  In weak coupling\footnote{There are both gauge and scalar fields,
but distributions they produced in zeroth order are the same.} the correction to $T_{00}$ is $1-2(y_1r)/L^2$ while
our strong coupling result gives 
\be {T_{00}\over T_{00}^{LO}}=1-{(y_1 r)\over z_m^2}\approx 1-0.34
    {2(y_1 r)\over L^2}\ee
The sign and the structure of the local field is the same,
while the magnitude is additionally reduced by about a factor 1/3.
What we learn from this comparison, once again,
 is that although a strongly coupled vacuum of the theory works
as a polarizable dielectric qualitatively, this is not true
literally.

\subsection{Is there a visible trace of the string?}
Another interesting question is the transverse distribution
of energy. In particular we calculate the r.m.s.:
$\sqrt{<y_2^2>}=
\(\frac{\int T_{00}y_2^2 dy_2}{\int T_{00}dy_2}\vert_{y_1=y_3=0}\)^{1\over2}$,
which characterizes the transverse energy distribution
on the middle plane between the quark-antiquark pair.

\ba
&&T_{00}\sim Q_{00}={1\over4}\int_{\infty}^0
dz\int \frac{s_{00}(z,\vx)P_s(\vy-\vx)}{z}d^3x \\
&&s_{00}(z<z_m)\sim \frac{E_1(z)}{3}+\frac{E_2(z)}{3}-\frac{F'(z)}{6z}  \no
&&s_{00}(z>z_m)\sim -\frac{G(z_m)}{6}  \nonumber
\ea

Note y-dependence enters only through the propagator $P_s$, we can
do the $y_2$ integral with the propagator first, then convolute
the result with the source $s_{00}$. The rest of the calculation
is straight forward. We will skip the details and only give the result:
$\sqrt{<y_2^2>}\approx0.41z_m$, while half the size of the dipole
is ${L\over2}\approx0.60z_m$. The r.m.s. is about ${1\over3}$ of the dipole
size, smaller than the perturbative result ${\sqrt{<y_2^2>}={L\over2}}$.

In order to make the trace of string clear, we would like to
rewrite (\ref{eq:dp_st2}) in a more physical form. This is done
by defining: $s^h_{mn}=s_{mn}-S_{mn}$, then we have
\be\label{eq:dp_pic}
&&Q_{mn}=-{1\over4}\int_0^\infty dz
\int \frac{S_{mn}(z,\vx)P_s(\vy-\vx)}{z}d^3x \no
&&-{1\over4}\int_0^\infty dz\int \frac{s^h_{mn}(z,\vx)P_s(\vy-\vx)}{z}d^3x
\ee

\begin{figure}[t]
  \epsfig{file=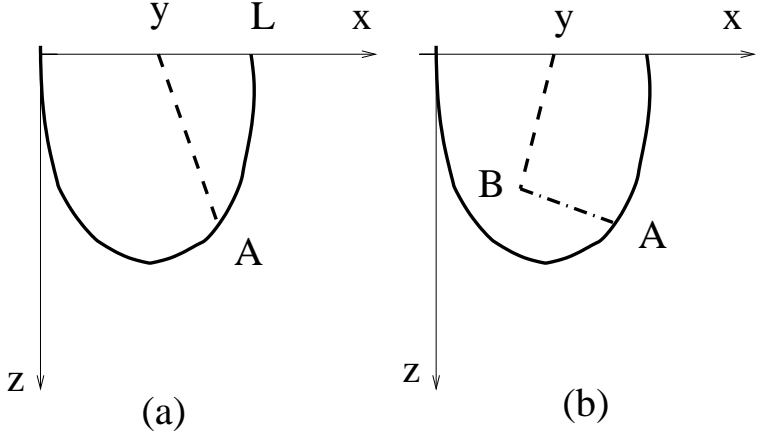,width=10.cm}
  \vspace{0.1in}
\caption{(color online) Schematic demonstration of the pending string
and the propagators of stress tensor. The source is at
the point $A$ integrated
over string, it either (a) goes directly to the observation point $y$
via bulk-to-boundary propagates(dashed line) , or (b) first transforms to 
$s^h_{mn}$ in some other point $B$ via bulk-to-bulk propagator(dash-dotted line),
then goes to the observation point} \label{fig_diagrams}
\end{figure}

The first piece is sourced by the original string $S_{mn}$, while the second
piece corresponds to contribution from $s^h_{mn}$. Since the latter
is obtained from $S_{\mu\nu}$ via (\ref{eq:h}) and (\ref{eq:hm}).
 The transform
from $S_{\mu\nu}$ to $s^h_{mn}$ can be interpreted 
as a bulk-to-bulk propagator, which is then attached to the 
bulk-to-boundary propagator $P_s$ to contribute to the stress tensor.
We schematically illustrate the two contributions in Fig.\ref{fig_diagrams}

We use the component $T_{00}$ as an example to study the relative
contribution from the two pieces:

\ba\label{eq:compare}
&&Q_{00}^1=-{1\over4}\int_0^\infty dz
\int \frac{S_{mn}(z,\vx)P_s(\vy-\vx)}{z}d^3x \no
&&={-1\over6}\int_0^{z_m}\frac{z\(\frac{z^5}{z_m^2\sqrt{z_m^4-z^4}}
+\frac{z\sqrt{z_m^4-z^4}}{z_m^2}\)}{(z^2+x(z)^2)^{7\over2}} \\
&&Q_{00}^2=-{1\over4}\int_0^{\infty} dz\int \frac{s^h_{mn}(z,\vx)P_s(\vy-\vx)}{z}d^3x \no
&&={-1\over6}\int_0^{\infty} dz\int_0^{x_m} dx^1\frac{z_m^4-3z_1^4}{z_1^2}
\theta(z-z_1)\frac{z}{(z^2+(x^1)^2)^{7\over2}} \no
&&={-1\over6}\int_0^{z_m}\frac{z_m^4-3z_1^4}{\sqrt{z_m^4-z_1^4}z_m^2}
\frac{1}{5(z_1^2+x^1(z_1)^2)^{5\over2}}
\ea

\begin{figure}[t]
  \epsfig{file=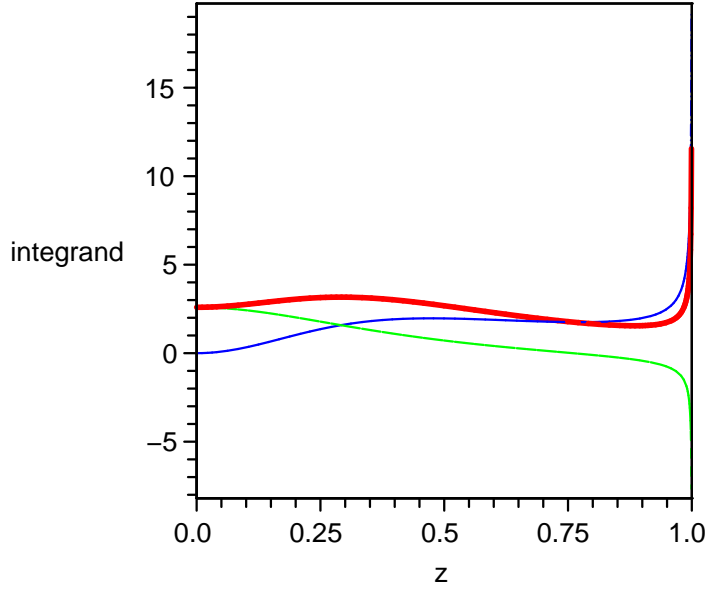,width=12.cm}
  \vspace{0.1in}
\caption{(color online) The
 integrands of the $z$ integral along the string for 
$Q_{00}^1$(blue dotted),$Q_{00}^2$
(green dashed) and their sum(red solid), with $z_m=1$} \label{fig_z_plot}
\end{figure}

We plot the integrands of (\ref{eq:compare}) in Fig.\ref{fig_z_plot}.
All three curves have a peak at $z=z_m$, which is due to geometry
of the string. However the peaks
are square root singularities of geometric
origin, which do not contribute significantly
to the integral and the finial $T_{00}$. Instead the latter receives significant contribution
from integration of all values of z. 

\section{A field of electric-magnetic dipole}
It is also interesting to consider the stress tensor of an quark and
monopole, in which case both electric and magnetic fields are
obviously present.
The string profile of the electric-magnetic dipole is obtained by
Minahan\cite{Minahan:1998xb}. It consists of
 a $(1,0)$ and a $(0,1)$ string, attached
to the quark and monopole at $z=0$ respectively, and a $(1,1)$ string
extending from $z=\infty$. The three string attach to each other
 at $z=z_0$, forming a Y-junction. With a suitable
choice of coordinate, we can
describe the $(1,1)$ string by $x^1=0$, and describe
the $(1,0)$ string and $(0,1)$ string profile
by $x^1=x(z_{m1},z)>0$ and $x^1=-x(z_{m2},z)<0$, 
where $z_{m1},z_{m2}$ are parameters of the string profile
$x(z)$. $x(z_m,z)$ satisfies $x_z=-\frac{z^2}{\sqrt{z_m^4-z^4}}$.
The parameters given by \cite{Minahan:1998xb} are:

\ba\label{eq:zm}
&&z_{m1}=z_0\alpha_1 \\
&&z_{m2}=z_0\alpha_2 \\
&&\alpha_1=\(\frac{1+t^2}{t^2}\)^{1\over4} \; \alpha_2=\(1+t^2\)^{1\over4} \nonumber
\ea
where $t={1\over g}$, g is the string coupling.

The action of a $(p,q)$ string is given by:
\be
&&S_{NG}=-\frac{\sqrt{p^2+q^2t^2}}{2\pi\alpha'}\int d^2\sigma\sqrt{-det g}
\ee

The $\dlt S_{\mu\nu}$ following from the action is:
\ba
&&\dlt S_{\mu\nu}(z<z_0)=\frac{-\ka^2 z}{2\pi\alpha'}
\dlt\(x^1-x(z_{m1},z)\)\dlt\(x^2\)\dlt\(x^3\)
\frac{1}{\sqrt{1+x_z^2}}\no
&&\begin{pmatrix}
{{x_z(z_{m1},z)^2+1}\over 3}& & & & \\
& {{2x_z(z_{m1},z)^2-1}\over 3}& -x_z(z_{m1},z)& & \\
& -x_z(z_{m1},z)& {{-x_z(z_{m1},z)^2+2}\over 3}& & \\
& & & {{2x_z(z_{m1},z)^2+2}\over3}& \\
& & & & {{2x_z(z_{m1},z)^2+2}\over3}
\end{pmatrix} \no
&&+\(x(z_{m1},z)\rightarrow -x(z_{m2},z)\)t  \;  \no
&&\dlt S_{\mu\nu}(z>z_0)=\frac{-\ka^2 z}{2\pi\alpha'}\dlt\(x^1\)\dlt\(x^2\)\dlt\(x^3\)
\begin{pmatrix}
{1\over 3}& & & & \\
& -{1\over 3}& & & \\
& & {2\over 3}& & \\
& & & {2\over3}& \\
& & & & {2\over3}
\end{pmatrix}\sqrt{1+t^2} \; 
\ea

We are not going to elaborate the calculation in any detail, as
the same
procedures for the electric dipole's case apply.
The far-field answer is
\be\label{eq:qm_far}
T_{mn}=\frac{-\sqrt{\lam}}{\pi}\frac{1}{8\pi}
\[-{2\over{3\lv y\rv^4}}
\begin{pmatrix}
1& & & \\
& 1& & \\
& & 1& \\
& & & 1
\end{pmatrix}+
\begin{pmatrix}
0& & & \\
& & & \\
& & y_my_n& \\
& & & 
\end{pmatrix}{4\over {3\lv y\rv^6}}
\]\sqrt{1+t^2} \label{eqn_source_em}
\ee
 Different from the case of electric dipole, in which
the leading total charge term drops out, the leading order
now is $1\over \lv y\rv^4$. The result is proportional
to  $\sqrt{\lambda(1+t^2)}=\sqrt{N({g_{YM}^2}+{(4\pi)^2/g_{YM}^2)}}$,
in good agreement with electric-magnetic duality\footnote{We remind
 the reader that Dirac condition in this theory is simply
that magnetic charge is the inverse of the electric one.}
 of the problem.  This shows the
electric-magnetic dipole looks like a dyon to distant observer.
 
Perturbatively one expect no correlation between electric and magnetic
charges, and the answer proportional to a sum
 $N(g_{YM}^2/4\pi+4\pi/g_{YM}^2)$, $without$ a square
 root. The reason a common square root appears
can again be traced to color correlation 
time by Shuryak and Zahed:
for example they have also shown that Coulomb, spin-spin
and spin-orbit forces are also united into one common
square root \cite{Shuryak:2003xa}.

For the near field, we recall the calculation of the previous section.
the LO stress tensor near the quark(monopole) is again the 
same as that of a single quark(monopole).
the NLO stress tensor only depend on the profile of the string
attached to the quark(monopole). Therefore, we can obtain the
stress tensor by the substitution: $z_m\rightarrow z_{m1}$(quark),
$z_m\rightarrow z_{m2}$(monopole). We display the NLO
near field result for the quark and monopole
 in (\ref{eq:q_near}),(\ref{eq:m_near}).

\be\label{eq:q_near}
&&T_{mn}=\frac{-\sqrt{\lam}}{\pi}\frac{1}{12\pi}\frac{1}{z_{m1}^2}
\biggl[
{y_1\over{6r_0^3}}
\begin{pmatrix}
5& & & \\
& 8& & \\
& & 8& \\
& & & 8
\end{pmatrix}-
\begin{pmatrix}
0& & & \\
& 2y_1& y_2& y_3\\
& y_2& & \\
& y_3& & 
\end{pmatrix}\frac{4}{3r_0^3} \no
&&-\begin{pmatrix}
0& & & \\
& & & \\
& & y_my_n& \\
& & & 
\end{pmatrix}\frac{y_1}{2r_0^5}
\biggr]
\ee

\be\label{eq:m_near}
&&T_{mn}=\frac{-\sqrt{\lam}}{\pi}\frac{1}{12\pi}\frac{t}{z_{m2}^2}
\biggl[
-{y_1\over{6r_0^3}}
\begin{pmatrix}
5& & & \\
& 8& & \\
& & 8& \\
& & & 8
\end{pmatrix}+
\begin{pmatrix}
0& & & \\
& 2y_1& y_2& y_3\\
& y_2& & \\
& y_3& & 
\end{pmatrix}\frac{4}{3r_0^3} \no
&&+\begin{pmatrix}
0& & & \\
& & & \\
& & y_my_n& \\
& & & 
\end{pmatrix}\frac{y_1}{2r_0^5}
\biggr]
\ee

The result at NLO suggests the impact of 
a monopole to the quark is the same as an antiquark
 at some distance  away. The precise relation
between the quark-monopole distance $L_{QM}$ and quark-antiquark
distance $L_{Q\bar Q}$ can be estimated. $L_{Q\bar Q}$ should be
chosen such that $z_{m}$ reproduce $z_{m1}$ for $L_{QM}$.
(2.6) and (3.2) of \cite{Minahan:1998xb} gives:

\be
&&L_{Q\bar Q}=2z_m\int_1^{\infty}\frac{dy}{y^2\sqrt{y^4-1}} \no
&&L_{QM}=z_0\(\alpha_1\int_{\alpha_1}^{\infty}\frac{dy}{y^2\sqrt{y^4-1}}
+\alpha_2\int_{\alpha_2}^{\infty}\frac{dy}{y^2\sqrt{y^4-1}}\) \no
&&\approx z_0\alpha_1\int_1^{\infty}\frac{dy}{y^2\sqrt{y^4-1}} \no
&&={1\over 2}L_{Q\bar Q}
\ee

where the approximation is due to the limit $g\rightarrow 0,\;
t\rightarrow\infty $. The result shows that in the above limit,
the quark feels the monopole like an antiquark at twice the distance.
Similarly, the monopole feels the quark at a distance $L$ like
an antimonopole $\frac{2\alpha_2}{\alpha_1}L=\frac{2}{\sqrt{g}}L$ away.

 Finally, let us address the ussue of the angular momentum
and Poynting vector. Perturbative
charge-monopole pair has at a generic point
  electric and magnetic
fields crossing at some angle,  thus producing a nonzero
 Poynting vector $T_{0m}\not= 0$. In fact its direction is
rotating around the line connecting charges, leading to nonzero
angular momentum of the field. In fact the
Dirac quantization condition is known to be directly
related to quantization of this angular momentum.

However, in our setting with Minahan's solution
 this effect is entirely absent and
there is no anular momentum or Poynting vector, $T_{0m}=0$. This can be traced
directly to the expression (\ref{eqn_source_em}) for the source
which has no such component. In gravity setting the energy-momentum
 of the Minahan
string construction does not care about direction of the magnetic
flux,
and the problem is again static and 
t-reflection symmetric. 

Perhaps the way to remedy the situation is to start with 
a different classical 
$rotating$ string, with some nonzero angular momentum, which
value is to be tuned to fit the Dirac
condition. If we will be able to make progress along this line,
we will report it elswhere\footnote{We thank Andrei Parnachev and
  Jinfeng Liao for helpful discussions of this issue. }.

\section{Summary and outlook}

The main results of this work are general expressions
for the stress tensor induced by objects in the AdS bulk 
(\ref{eq:sq_st}),(\ref{eq:ff}),(\ref{eq:nf1}),(\ref{eq:nf2}),
(\ref{eq:qm_far}),(\ref{eq:q_near}),(\ref{eq:m_near}).
 In general, we found that
two components of gravity perturbation -- the trace of the
metric $h$ and its tensor part $h_{\mu\nu}$ -- have different
equations and Green functions. Although
$h$ itself on the boundary does not have $O(z^4)$ corrections
or induced stress tensor (as follows from conformal symmetry of the
boundary theory),  two components are intermixed in curved background
and thus $h$ (incorporated into a ``generalized source'')
leads to physical effects including the stress tensor. 

 General formulae
 are then used for static electric and electric-magnetic
dipoles, as important examples.
Confidence in the results come from checking all of them for
tracelessness and energy-momentum conservation.  We worked out
the far field asymptotic, as well as an expressions for the
 field near one of the charges. 

The far distance asymptotic of the stress tensor is $\sim L^3/r^7$,
the same as in previous calculation \cite{Callan:1999ki} for
dim-4 scalar density, the angular distribution is different.  
We found that although all angular
 structures are as expected from perturbative analysis for dipoles,
the coefficients (and angular distribution of stress tensor)
are quite different
from the weak coupling limit. The same is found for the
near-field domain. It means although
a naive idea of strongly coupled
vacuum acting as a dielectric qualitatively is holding, quantitatively 
 it definitely fails.  

We also found that on the boundary
there seems to be no visible trace of a string.  In fact
even in between the two charges (e.g. at $y_1=0$) the dominant
contribution still comes from ``vertical'' parts of the string   
rather than its ``horizontal'' part directly beneath 
the observation point. The distribution looks like two distorted
polarization clouds about two charges, instead of a string-like
object. 

This conclusion is relevant for interpretation of string-like
entity which seems to appear
via linear part in static dipole potentials
on the lattice at $T$ just above deconfinement for QCD-like theories.
We think those are
due to some flux tubes. Their formation
is due to phenomena which needs more specific ingredients than
just a strong coupling regime.
Let us further conjecture that the distributions
we calculated from AdS/CFT should instead be  similar to those
in QCD-like theories in a ``quasi-conformal regime'', at
temperatures not too close to deconfinement, $T>1.5Tc$.
This is the region in which flux tube effects are gone, the
potentials become a screened-Coulomb type and thermodynamical
observables are about constant when divided by appropriate
powers of $T$. This conjecture will be tested directly 
in forthcoming
lattice calculations.

As an outlook for this work we have in mind, we would like
 to work out stress tensor imprints of dynamical (rather than static)
objects. In particular, those are ``debris'' created in 
high energy heavy ion collisions, see \cite{SSZ} for a basic
picture and to our previous paper \cite{Lin:2006rf} in which we
formulated the picture and calculated
trajectories of different types of objects falling into AdS bulk.
We hope then elucidate the process of black hole formation,
out of those ``debris'' and see whether the stress tensor
imprints would be approaching hydrodynamical solutions, which were
so successful for the description  \cite{SZ12} of RHIC data.

\noindent{\large \bf Acknowledgments} \vskip .35cm We thank
I.Zahed and S.-J.Sin for multiple discussions. Our
 work was partially
supported by the US-DOE grants DE-FG02-88ER40388 and
DE-FG03-97ER4014.

\appendix
\section{Linearization of Ricci tensor}\label{app:Rmunu}

We may start with the following relations:
\ba
&&\dlt R_{\mu\nu}=\dlt\Gamma_{\mu\lam;\nu}^{\lam}-
\dlt\Gamma_{\mu\nu;\lam}^{\lam} \no
&&\dlt\Gamma_{\mu\nu}^{\lam}={1\over2}g^{\lam\sigma}
(\dlt g_{\sigma\mu;\nu}+\dlt g_{\sigma\nu;\mu}-\dlt g_{\mu\nu;\sigma})\nonumber
\ea

Since the covariant derivative on the metric vanishes, $g^{\lam\sigma}_{;\mu}=0$,
the metric commutes with the covariant derivative. $\dlt R_{\mu\nu}$
can be further simplified.
\be
&&\dlt R_{\mu\nu}={1\over2}g^{\lam\sigma}
(-h_{\sigma\mu;\nu;\lam}-h_{\sigma\nu;\mu;\lam}+h_{\sigma\lam;\mu;\nu}
+h_{\mu\nu;\sigma;\lam}) \no
&&=-{1\over2}g^{\lam\sigma}(h_{\sigma\mu;\nu;\lam}+h_{\sigma\nu;\mu;\lam})
+{1\over2}h_{;\mu;\nu}+{1\over2}g^{\lam\sigma}h_{\mu\nu;\sigma;\lam}
\ee
with $h_{\mu\nu}=\dlt g_{\mu\nu} \;h=g^{\lam\sigma}h_{\lam\sigma}$.

We choose to work in Poincare coordinate, the only nonvanishing
Christoffels of which are:
\be
\Gamma_{tt}^z=\Gamma_{zz}^z=-{1\over z} ,\; \Gamma_{x^ix^i}^z={1\over z},
\; \Gamma_{tz}^t=\Gamma_{x^iz}^{x^i}=-{1\over z}
\ee
We calculate the components $\dlt R_{zz}$,$\dlt R_{zm}$,$\dlt R_{mn}$
separately. Through tedious algebra, we arrive at:
\ba
&&\dlt R_{zz}={1\over 2}h_{,z,z}-{1\over {2z}}h_{,z} \\
&&\dlt R_{zm}={1\over 2}\(h_{,m}-h_m\)_{,z} \\
&&\dlt R_{mn}={1\over 2}\square h_{mn}+2h_{mn}+{z\over 2}h_{mn,z}
-{1\over 2}\(h_{m,n}+h_{n,m}\)+{1\over 2}\(h_{,m,n}-\Gamma_{mn}^z
h_{,z}\)
\ea
with $h_m=g^{\lam\sigma}h_{\lam m,\sigma}$

\vskip 1cm

\end{document}